\documentclass[11pt,a4paper]{article}

\usepackage{booktabs}
\usepackage{siunitx}
\usepackage{amsmath}
\usepackage[margin=2.5cm]{geometry}
\usepackage{graphicx}
\usepackage{xcolor}
\usepackage{url}
\usepackage[hidelinks]{hyperref}

\newcommand{\archivefield}[2]{%
  \section*{#1}%
  #2%
}

\newcommand{\funding}[1]{\archivefield{Funding}{#1}}
\newcommand{\roles}[1]{\archivefield{Author contributions}{#1}}
\newcommand{\data}[1]{\archivefield{Data availability}{#1}}
\newcommand{\suppdata}[1]{\archivefield{Supplementary material}{#1}}
\newcommand{\restrictedzenodolink}{https://zenodo.org/records/20734774?preview=1\&token=eyJhbGciOiJIUzUxMiJ9.eyJpZCI6ImJhYjhkMDU2LTI3ZjUtNDA3MS05NGM0LTViZTc1OTUwYWM4MSIsImRhdGEiOnt9LCJyYW5kb20iOiJlMjllMjkzZDY2ODY1NjVlNmYxMWQwNzU3NWE2NDIyNCJ9.vJLOA4kyBXJ8WlAL_j5ewevLIsKfUCf_vxK36opE7_ZNpqJefaxFpoEZGsnZEAaMQeUPOJ-37GrQZBFlMssqNQ}

\definecolor{phantomBone}{HTML}{9C7A4B}
\definecolor{phantomBrain}{HTML}{5B8CC0}
\definecolor{phantomSkin}{HTML}{E3B08D}
\definecolor{phantomAir}{HTML}{F4F4F4}
\definecolor{phantomGTV}{HTML}{D64B4B}
\newcommand{\phantomlegendbox}[1]{\begingroup\fboxsep=0pt\colorbox{#1}{\rule{0pt}{0.8em}\hspace{1.1em}}\endgroup}

\begin{document}

\title{OpenPINT---Open-source Planning for Isoeffective Nuclear Treatments in BNCT research}

\author{%
Ian Postuma\textsuperscript{1,*} \\
Sara J. Gonz\'alez\textsuperscript{2,3,4} \\
Setareh Fatemi\textsuperscript{1} \\
Cristina Pezzi\textsuperscript{5,1} \\
Carolina Ruzzon\textsuperscript{5} \\
Oreste Nicrosini\textsuperscript{1} \\
Valerio Vercesi\textsuperscript{1} \\
Silva Bortolussi\textsuperscript{5,1} \\
\\
\textsuperscript{1}Istituto Nazionale di Fisica Nucleare, unit of Pavia, Italy \\
\textsuperscript{2}Comisi\'on Nacional de Energ\'ia At\'omica, CNEA, Buenos Aires, Argentina \\
\textsuperscript{3}Consejo Nacional de Investigaciones Cient\'ificas y T\'ecnicas, CONICET, Buenos Aires, Argentina \\
\textsuperscript{4}Universidad Nacional de San Mart\'in, Buenos Aires, Argentina \\
\textsuperscript{5}Università degli studi di Pavia, Dipartimento di Fisica ``Alessandro Volta'', Pavia, Italy \\
\textsuperscript{*}Author to whom any correspondence should be addressed: \url{ian.postuma@pv.infn.it}
}

\date{Preprint manuscript}

\maketitle

\begin{abstract}
\textbf{Objective:} Present \texttt{OpenPINT} (Open-source Planning for Isoeffective Nuclear Treatments), an open-source treatment planning system for nuclear therapies that integrates Monte Carlo dose calculations with modular dosimetric and radiobiological models for \allowbreak photon-\allowbreak isoeffective dose evaluation.

\textbf{Approach:} We describe the software architecture, implementation choices, and data flow from segmented geometry and source configuration to NIfTI dose outputs. We define BNCT-relevant dosimetric metrics and evaluate the workflow with reproducible analytic and voxelized cylindrical-phantom benchmarks, supplemented by a geometric patient-positioning example.

\textbf{Main results:} The module provides a reproducible and scriptable path for generating MCNP-ready inputs, extracting component-wise BNCT dose maps, and computing analysis-ready outputs for quality checks and decision support. Fine-resolution voxelized configurations reproduced the 1 mm analytic reference within 0.13\% for the brain-limited irradiation-time endpoint, whereas the full voxelized sweep exposed deviations up to 4.42\% in coarse 8--10 mm configurations. Patient-wide gamma pass rates were at least 99.60\% for the evaluated mesh/interpolation cases, while low-dose DVH-tail quantities remained sensitive to boundary discretization.

\textbf{Significance:} This first paper isolates and validates the simulation-preparation and dosimetric-analysis core of an open-source BNCT treatment-planning platform. It establishes a foundation for subsequent work on optimization, biological weighting, and clinical workflow integration.
\end{abstract}


\section*{Preprint notice}
This manuscript is a preprint released before peer review. It has not been peer reviewed and should not be interpreted as evidence of clinical suitability. It is provided to document the OpenPINT methodology, benchmark data, and software state at the time of public dissemination. The software and accompanying data are intended for research, education, reproducibility, and method-development use only.

\section{Introduction}

Boron neutron capture therapy (BNCT) combines neutron transport physics, patient-specific geometry, boron pharmacokinetics, and radiobiological interpretation in a workflow that is technically demanding and highly sensitive to implementation details. The current international state of practice emphasizes the need for reproducible planning pipelines, explicit uncertainty handling, and software traceability across simulation and dosimetric analysis stages~\cite{iaea2023bnct}.

Treatment planning systems are central to this process because they translate anatomical images, material assignments, beam geometry, and biological weighting assumptions into predicted absorbed and weighted dose distributions. In conventional photon, proton, and ion therapy, open research platforms such as matRad, SlicerRT, and OpenTPS have shown that openly available planning software can support education, benchmarking, and algorithm development without replacing commissioned clinical treatment planning systems~\cite{wieser2017matrad,pinter2012slicerrt,wuyckens2023opentps}. These projects are useful precedents because they make a clear distinction between research software and clinically approved medical devices.

BNCT has its own TPS history. Dedicated systems have been developed around reactor- and accelerator-based facilities, including NCTPLAN, SERA, THORplan, JCDS/Tsukuba-Plan, NeuCure Dose Engine, and more recent accelerator-BNCT planning environments such as NeuMANTA~\cite{zamenhof1996nctplan,nigg1999sera,lin2011thorplan,kumada2018tro,kumada2018multimodal,hu2021neucure,hu2025application}. These systems demonstrate the essential BNCT-specific requirements: conversion of medical images into transport-ready patient models, coupled neutron/photon Monte Carlo transport, component-wise dose reconstruction, biological weighting, and reporting of structure-based planning metrics. However, much of the BNCT TPS ecosystem has historically been facility-specific, closed, or linked to particular clinical devices and local beam models.

In BNCT treatment-planning research, the computational chain is therefore often fragmented: geometry preparation, Monte Carlo input authoring, simulation execution, mesh parsing, spatial remapping, and metric extraction may rely on separate scripts and manual file manipulation. This fragmentation increases the risk of silent errors, including coordinate mismatches, label inconsistencies, and undocumented preprocessing choices, and it limits reproducibility and inter-comparison across institutions.

For open BNCT treatment-planning research, a dedicated software platform should therefore: (i) formalize the Monte Carlo input preparation process, (ii) standardize output parsing into analysis-ready medical image formats, (iii) preserve component-wise dose information before biological weighting, (iv) expose dosimetric quantities relevant to planning decisions, and (v) support regression testing on controlled geometries before patient-oriented studies.

\subsection{Scope of this first computational release}
This paper presents the initial open-source release of \texttt{OpenPINT}, Open-source Planning for Isoeffective Nuclear Treatments, as a software/methodology contribution focused on two tightly coupled tasks: MCNP simulation preparation and dosimetric output analysis. The repository is intended to grow into a broader treatment-planning platform for nuclear therapies, while the present manuscript deliberately validates only the simulation-preparation and dose-analysis core as an independently testable and citable foundation.

The software described here is not a clinically certified TPS and is not intended for direct patient treatment decisions without independent commissioning, local beam-model validation, quality assurance, and regulatory review. Its role is analogous to open research TPS platforms in other radiotherapy domains: it provides transparent and reproducible infrastructure for testing algorithms, comparing discretization choices, archiving benchmark cases, and developing BNCT planning methods. The scope intentionally excludes full treatment optimization loops, advanced biological model calibration, and complete clinical decision workflow orchestration. These broader components will be addressed in later TPS-focused work, involving biophysical modelling considerations, prescribing strategies, and dose-reporting methods. This separation keeps the present manuscript centered on computational correctness, software architecture, and reproducible dosimetric data flow, while making the codebase available as an extensible open-source platform.

\section{Methods}
\subsection{Software architecture}
The implementation is organized as an open-source Python research codebase with both API-level and command-line access. The Monte Carlo neutron/photon transport code used in the calculations presented here is MCNP~\cite{MCNP}. Other codes such as PHITS~\cite{PHITS} or Geant4~\cite{geant4} could be considered in future by changing the input syntax and output-reading layer. The \texttt{OpenPINT.mcgenerator} package handles geometry/mask preparation and MCNP input construction; \texttt{OpenPINT.dose} handles mesh reading, dose-component reconstruction, and planning-oriented dosimetric analysis; and \texttt{bin/} provides workflow-oriented CLI entry points for reproducible execution.

From an architectural perspective, the pipeline is divided into three layers:
\begin{enumerate}
\item \textit{Input preparation layer}: NIfTI~\cite{NIfTI} masks, material definitions, and simulation settings are transformed into MCNP-ready textual inputs.
\item \textit{Transport-output decoding layer}: MCNP mesh tallies are parsed and mapped to structured component dose arrays.
\item \textit{Analysis layer}: component maps are exported to CT-aligned NIfTI, combined across fractions, and summarized through planning-relevant metrics.
\end{enumerate}

This separation allows each layer to be validated independently and makes interfaces explicit (file schema, coordinate mapping, and component naming).

\subsection{MCNP input preparation workflow}
The MCNP preparation workflow starts from segmented NIfTI images. In the current release, the input-generation layer accepts structure masks such as brain, bone/skull, skin, and gross tumour volume (GTV), together with a user-defined ordering of material classes. These masks define the patient or phantom materials to be transported in MCNP. Voxels outside the explicitly segmented anatomy are assigned to an AIR class by taking the complement of the supplied masks, preventing unlabeled voxels from entering the simulation geometry.

The mask set is collapsed into a single simulation mask in which integer labels follow the selected material order. This label image is saved as a NIfTI object for inspection and is also converted to an MCNP lattice. During lattice generation, the image affine is used to preserve anatomical orientation: voxel traversal is reversed along axes with negative affine direction cosines so that the MCNP fill order remains consistent with the medical-image coordinate convention. The resulting lattice is therefore not an arbitrary array dump, but a deterministic translation from segmented NIfTI space to MCNP voxel geometry.

Material properties are kept separate from the segmentation masks in a JSON material library. For each material class, the library specifies the MCNP material identifier, mass density, isotope composition through ZAID/fraction pairs, and optional thermal treatment information. This design allows the same segmented anatomy to be re-rendered with alternative material assumptions without changing the mask data. Material assignment is generated in two coordinated MCNP blocks: (i) cell cards associating lattice labels with material identifiers and densities, and (ii) material cards containing the isotope composition and thermal scattering definitions. A template-based renderer then fills placeholders in an MCNP input skeleton to produce simulation-ready files.

Simulation geometry extents and voxel box boundaries are computed from image spacing, shape, affine translation, and irradiation reference point. Coordinates are converted to MCNP units (cm), and lattice dimensions are propagated consistently to mesh-tally definitions. Input checks include positive dimension constraints, geometry-size consistency, and parameter-range validation in the CLI workflow.

The material definitions used for the cylindrical validation study are reported in Table~\ref{tab:phantom-materials}. MCNP negative fractions denote mass fractions. Brain, skin, and GTV were treated as hydrogenous soft-tissue materials and assigned the MCNP light-water thermal scattering treatment (\texttt{lwtr.01}), following the water thermal-treatment convention used in the NCTPlan/MacNCTPlan BNCT planning lineage~\cite{kiger1996macnctplan,zamenhof1996nctplan}. Bone and air were not assigned a thermal scattering treatment in the current benchmark material library.

\begin{table}[htbp]
\caption{MCNP material definitions used in the cylindrical phantom benchmark.}
\label{tab:phantom-materials}
\centering
\scriptsize
\setlength{\tabcolsep}{3pt}
\begin{tabular}{p{0.12\linewidth}p{0.78\linewidth}}
\toprule
Class & MCNP material definition \\
\midrule
BONE & $\rho=\SI{1.610}{g.cm^{-3}}$; H 0.034, C 0.155, N 0.042, O 0.435, Na 0.001, Mg 0.002, P 0.103, S 0.002985, Ca 0.225, $^{10}$B 0.000015; no thermal treatment. \\
BRAIN & $\rho=\SI{1.040}{g.cm^{-3}}$; H 0.107, C 0.145, N 0.022, O 0.712, Na 0.002, P 0.004, S 0.002, Cl 0.003, K 0.002985, $^{10}$B 0.000015; \texttt{lwtr.01}. \\
SKIN & $\rho=\SI{1.090}{g.cm^{-3}}$; H 0.100, C 0.143, N 0.034, O 0.708, Na 0.002, P 0.003, S 0.005, Cl 0.002, K 0.002985, $^{10}$B 0.000015; \texttt{lwtr.01}. \\
AIR & $\rho=\SI{0.001205}{g.cm^{-3}}$; C 0.000124, N 0.755268, O 0.231781, Ar 0.012827; no thermal treatment. \\
GTV & $\rho=\SI{1.030}{g.cm^{-3}}$; H 0.107, C 0.145, N 0.022, O 0.712, Na 0.002, P 0.004, S 0.002, Cl 0.003, K 0.002960, $^{10}$B 0.000040; \texttt{lwtr.01}. \\
\bottomrule
\end{tabular}
\end{table}

\subsection{Synthetic cylindrical phantom generator}
\begin{figure}[htbp]
\centering
\includegraphics[width=\linewidth]{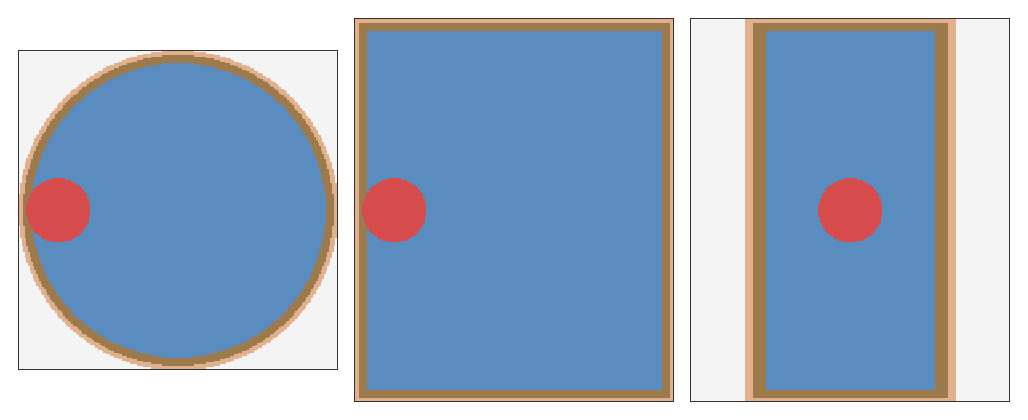}
\vspace{0.5em}
\begin{tabular}{ccccc}
\phantomlegendbox{phantomBone} \textcolor{phantomBone}{Skull/bone} &
\phantomlegendbox{phantomBrain} \textcolor{phantomBrain}{Brain} &
\phantomlegendbox{phantomSkin} \textcolor{phantomSkin}{Skin} &
\phantomlegendbox{phantomAir} Air &
\phantomlegendbox{phantomGTV} \textcolor{phantomGTV}{GTV}
\end{tabular}
\caption{Synthetic cylindrical phantom generated by the \texttt{OpenPINT} input-preparation workflow. The panels show orthogonal sections through the tumour centroid and preserve physical voxel aspect ratio; the axial panel spans \SI{200}{mm} by \SI{200}{mm}, whereas the coronal and sagittal panels span \SI{200}{mm} by \SI{240}{mm}.}
\label{fig:cylindrical-phantom}
\end{figure}

To support controlled validation independently of patient data, the library includes a cylindrical phantom generator, a geometry already used for BNCT research studies~\cite{Provenzano2019,postuma2021biology}. The generator creates a synthetic NIfTI geometry with a cylindrical external contour, concentric skin and skull shells, an internal brain region, and a spherical GTV placed at a user-defined depth from the external surface. The default configuration uses a \SI{100}{mm} outer radius, \SI{240}{mm} cylinder height, \SI{3}{mm} skin thickness, \SI{5}{mm} skull thickness, and a \SI{20}{mm} tumour radius with the tumour centre \SI{25}{mm} from the cylinder surface. Voxel spacing, lattice shape, bounding-box padding, tumour size, tumour depth, and FMESH interval counts can be varied from the command line (Fig. \ref{fig:cylindrical-phantom}).

The phantom generator exports both individual binary masks and the combined simulation mask used by the MCNP pipeline. The same mask-to-material logic described above is then applied with the material order BONE, BRAIN, SKIN, AIR, and GTV. In addition to the NIfTI masks, the workflow writes a bounding-box JSON file, a case-summary JSON file containing geometry and mesh metadata, and a complete MCNP input generated from the local beam-shaping assembly template (i.e. an input template file containing the BSA and beam port geometrical configuration). This makes the phantom suitable for discretization studies in which anatomical voxel size and tally mesh size are varied independently. The cylindrical-phantom simulations reported in this work used the neutron beam model developed by Postuma et al. for BNCT beam evaluation~\cite{postuma2021biology}. 

\subsection{Patient-positioning and beam-entry geometry}
\texttt{OpenPINT} also includes a geometric patient-positioning strategy for selecting an irradiation orientation before MCNP input generation. This strategy is demonstrated in the \texttt{examples/patient\_rotation/} workflow. The positioning example is a separate illustrative geometry from the cylindrical benchmark above: it uses the same \SI{100}{mm} patient radius and \SI{240}{mm} height, but a smaller spherical GTV with a \SI{10}{mm} radius, sampled on a \SI{2}{mm} voxel grid. The objective of the current implementation is not fluence or dose optimization; rather, it computes a reproducible rigid transform that places a selected patient-surface entry point at the beam port and aligns the entry-to-tumour direction with the beam axis.

The workflow first computes the GTV centroid from the segmented tumour mask in patient RAS millimetres. The centroid is defined as the arithmetic mean of the RAS coordinates of the GTV foreground voxel centres, and therefore represents the mask centre of mass under uniform voxel weighting. A point on the patient surface is then extracted either from an explicit skin mask or from the CT image using a Hounsfield-unit threshold followed by optional morphological cleanup and largest-component selection. When a CT-derived patient mask is used, the entry-point search is performed on the exposed boundary faces of the patient mask, returning a physical surface coordinate rather than only the centre of the nearest surface voxel.

\begin{figure}[htbp]
\centering
\includegraphics[width=\linewidth]{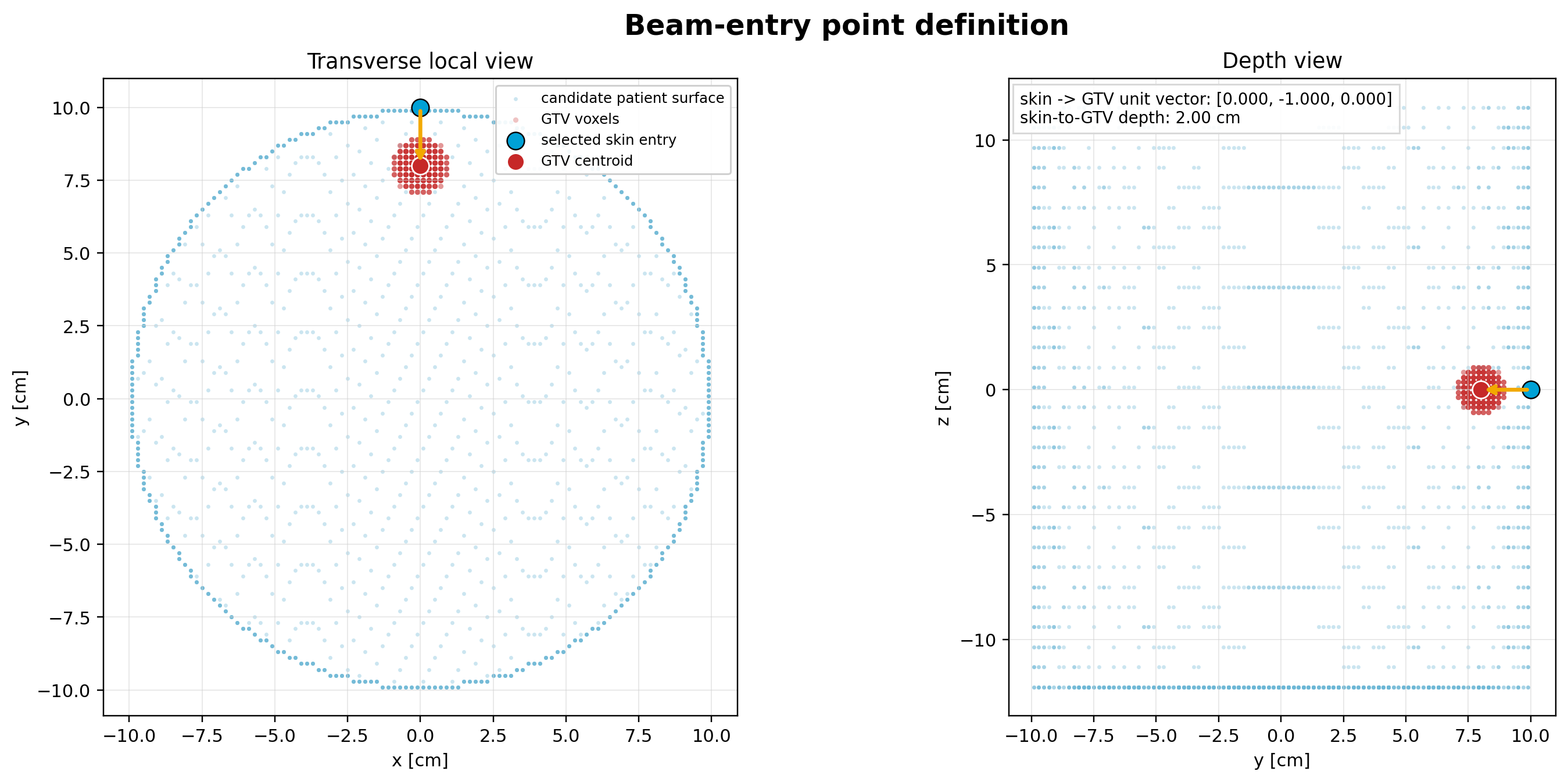}
\caption{Definition of the geometric points used for patient positioning in the \texttt{examples/patient\_rotation/} workflow. The generated beam-alignment image shows the candidate patient surface, GTV voxels, selected skin-entry point, GTV centroid, skin-to-GTV unit vector, and skin-to-GTV depth. The selected skin-entry point is the patient-surface point that minimizes the Euclidean distance to the GTV centroid.}
\label{fig:patient-rotation-points}
\end{figure}

The candidate skin-entry point is defined as the surface point that minimizes the Euclidean distance to the GTV centroid in RAS space. If $\mathbf{c}_{\mathrm{GTV}}$ is the GTV centroid and $\mathcal{S}$ is the set of candidate patient-surface points, the selected entry point is
\begin{equation}
\mathbf{s}_{\mathrm{entry}} =
\underset{\mathbf{s}\in\mathcal{S}}{\arg\min}\,
\left\|\mathbf{s}-\mathbf{c}_{\mathrm{GTV}}\right\|_2 .
\end{equation}
This provides an automatic geometric approximation of the shortest skin-to-target path and defines the entry-to-target vector $\mathbf{v}=\mathbf{c}_{\mathrm{GTV}}-\mathbf{s}_{\mathrm{entry}}$ (Figure~\ref{fig:patient-rotation-points}).

The selected tumour centroid and skin-entry point are converted from NIfTI RAS coordinates to a centred MCNP local coordinate frame in centimetres. The conversion requires a positive-diagonal image affine, so that voxel centres and physical surface coordinates can be mapped deterministically into MCNP local coordinates. A rigid rotation is then computed so that the entry-to-target vector is parallel to the requested beam axis. Optional Euler pre-rotations $(\alpha,\beta,\gamma)$ allow the user to impose a preferred patient roll or orientation convention before the corrective alignment is applied.

\begin{figure}[htbp]
\centering
\includegraphics[width=\linewidth]{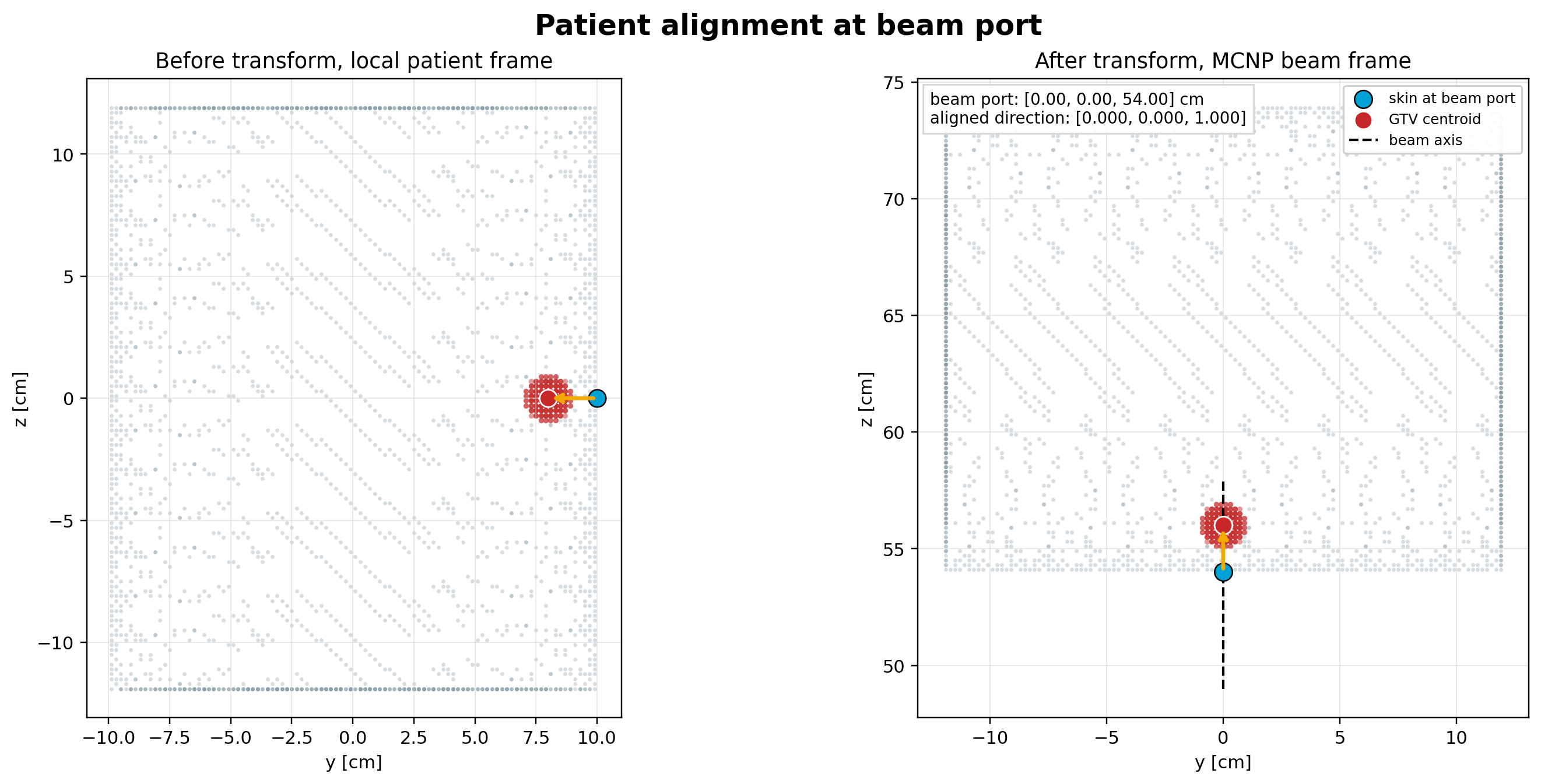}
\caption{Beam-port alignment after applying the computed rigid patient transform. This separate positioning example uses a spherical GTV of \SI{10}{mm} radius, compared with the \SI{20}{mm} radius GTV in Figure~\ref{fig:cylindrical-phantom}. The left panel shows the selected skin-entry point and GTV centroid before transformation in the local patient frame. The right panel shows the transformed geometry in the MCNP beam frame, where the skin-entry point is placed at the beam port and the entry-to-target direction is aligned with the requested $+z$ beam axis.}
\label{fig:patient-rotation-alignment}
\end{figure}

Finally, the translation vector is chosen so that the transformed skin-entry point coincides with the beam-port coordinate. In the example case, the beam axis is $+z$ and the beam port is located at $(0,0,54)$ cm. The resulting rotation and translation are written to the MCNP input as a \texttt{TR51} transformation card, replacing the placeholder transform in the beam-shaping assembly template. The case summary records the GTV centroid, selected skin-entry point, minimum skin-to-centroid distance, beam axis, beam port, rotation matrix, translation vector, transformed skin and tumour coordinates, and the final MCNP transform card. This makes the positioning step auditable and allows multiple candidate orientations to be generated and compared using the same downstream transport and dosimetric analysis pipeline. The transformed beam-port alignment is shown in Figure~\ref{fig:patient-rotation-alignment}.

\subsection{Output parsing and dosimetric analysis}
For MCNP outputs, the implemented BNCT dose components are obtained from four component-specific flux mesh tallies. Each tally scores particle fluence on the requested spatial mesh and is converted into a dose-rate contribution by multiplication with the corresponding energy-dependent kerma factors through the MCNP tally multiplier formalism. The \texttt{OpenPINT} parser then reads the resulting mesh-tally blocks and decodes them into 3D component arrays. The four component channels are handled with a fixed semantic convention: \texttt{B10}, \texttt{N14}, \texttt{n}, and \texttt{g}. Here, \texttt{B10} represents the boron dose contribution from the $^{10}$B(n,$\alpha$)$^7$Li reaction per unit concentration of $^{10}$B, \texttt{N14} represents the nitrogen dose contribution from the $^{14}$N(n,p)$^{14}$C reaction, \texttt{n} represents the non-capture neutron kerma contribution, and \texttt{g} represents the photon kerma contribution.

\begin{figure}[htbp]
\centering
\includegraphics[width=\linewidth]{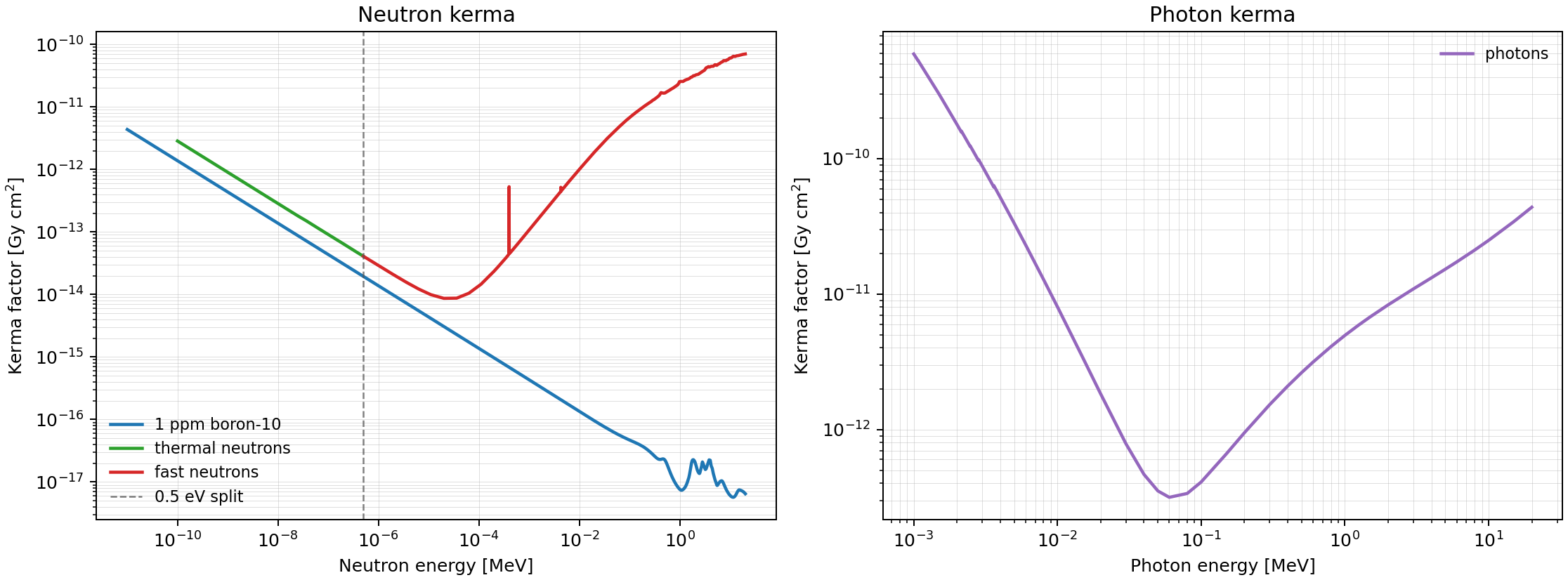}
\caption{Energy-dependent kerma factors used by the MCNP component tallies in the simulation template. The left panel shows the neutron kerma factors for \SI{1}{ppm} boron-10, thermal neutrons, and fast neutrons; the right panel shows the photon kerma factors. The dashed line marks the \SI{0.5}{eV} thermal/fast neutron split used to separate the thermal and fast neutron kerma channels.}
\label{fig:kerma-factors}
\end{figure}

The boron kerma factors used for the \texttt{B10} channel are tabulated for \SI{1}{ppm} of $^{10}$B in any material and were calculated from ENDF/B-VI cross sections and reaction Q-values as reported by Goorley et al.~\cite{goorley2002reference}. The neutron kerma factors used for the \texttt{N14} and \texttt{n} channels correspond to ICRU 44 brain material and were calculated from ICRU Report 63 kerma data with JENDL-3.2 cross sections and Q-values, again following the reference BNCT dosimetry data set of Goorley et al.~\cite{goorley2002reference}. The photon kerma factors used for the \texttt{g} channel correspond to ICRU 44 brain material and were calculated from the NIST photon mass attenuation and mass energy-absorption coefficient data of Hubbell and Seltzer~\cite{hubbell1995nistir5632}. The energy-dependent kerma factors used in the MCNP component tallies are shown in Figure~\ref{fig:kerma-factors}. In this implementation, the MCNP tally output is therefore already component-wise kerma-weighted; the Python output-processing layer preserves these four physical components before any boron-concentration scaling, biological weighting, summation, or DVH analysis is applied. This ensures flexibility and personalization of the biophysical models used to calculate biologically weighted doses, which is important for inter-comparison and reporting.

For PHITS outputs, an equivalent component-specific parser is used when the same semantic channels are available.

Component arrays are resampled and embedded into the original CT geometry using a simulation-region shape and voxel offset. CT affine and header metadata are reused to guarantee spatial compatibility with downstream planning tools. The resulting NIfTI outputs are written as component-wise maps, enabling direct visualization and ROI-based analyses.

When multiple irradiation sessions are planned, the workflow computes weighted component averages using per-fraction weights from a tabular input file. Optional time normalization scales dose-rate quantities to per-fraction or protocol-specific reference times.

\subsection{BNCT-specific planning logic and dosimetric metrics}
The module keeps a strict distinction between physical component doses and biologically weighted outputs. For a voxel $\mathbf{x}$ and irradiation time $t$, the physical absorbed dose can be expressed as

\begin{equation}
D_{\mathrm{phys}}(\mathbf{x}) =
t\left(C_B\dot{D}_{\mathrm{B10}}(\mathbf{x}) +
\dot{D}_{\mathrm{N14}}(\mathbf{x}) +
\dot{D}_{\mathrm{n}}(\mathbf{x}) +
\dot{D}_{\mathrm{g}}(\mathbf{x})\right),
\end{equation}

where $C_B$ is the boron concentration used for the selected tissue.

For weighted reporting, component-specific multipliers can be applied:
\begin{equation}
D_{\mathrm{w}}(\mathbf{x}) =
t\left(r_B C_B\dot{D}_{\mathrm{B10}}(\mathbf{x}) +
r_{N14}\dot{D}_{\mathrm{N14}}(\mathbf{x}) +
r_n\dot{D}_{\mathrm{n}}(\mathbf{x}) +
r_g\dot{D}_{\mathrm{g}}(\mathbf{x})\right).
\end{equation}

Here, $r_B$ is the compound biological effectiveness (CBE) factor assigned to the boron component, whereas $r_{N14}$, $r_n$, and $r_g$ are biological weighting factors, also referred to as relative biological effectiveness (RBE) factors, for the nitrogen-capture, non-capture neutron, and photon components, respectively~\cite{iaea2023bnct}.

Within this standalone module, these outputs support practical planning summaries such as limiting-OAR irradiation time estimates, component-wise dose inspection at critical points, region-wise descriptive statistics, and optional biologically-weighted dose exploratory reporting.

\subsection{Implemented radiobiological dose formalisms}
Although the present paper focuses on MCNP input generation and dosimetric-output processing, the TPS codebase also contains a model database for evaluating BNCT dose distributions with alternative dosimetric and radiobiological formalisms. These models are implemented in \texttt{OpenPINT.dose.Model\_db} and are exposed through the treatment-plan layer for research use.

The implemented functions include so far physical absorbed-dose summation, RBE/CBE-weighted dose \cite{Coderre1999TheTherapy}, photon isoeffective dose (IsoE)~\cite{gonzalez2012photon}, tumour control probability (TCP)~\cite{gonzalez2017photon}, normal-tissue complication probability (NTCP)~\cite{viegas2024predicting}, uncomplicated tumour control probability (UTCP)~\cite{Provenzano2019}, biologically effective dose (BED), and a dose-independent weighting approximation for normal-tissue photon IsoE~\cite{pedrosa2020isoe}. The weighted-dose database at this moment stores parameter sets for conventional CBE/RBE reporting, including tumour and healthy-tissue presets, and tissue-specific values compiled for brain, skin, liver, lung, kidney, bladder, and tumour in the context of BNCT beam evaluation~\cite{postuma2021biology}. The IsoE database includes parameter sets for in-vivo head-and-neck modelling ~\cite{gonzalez2017photon}, in-vitro glioblastoma and gliosarcoma models \cite{Marcaccio}, lip mucosa, and additional in-vitro parameterizations. TCP presets include and head-and-neck~\cite{monti2017boron} parameter sets, while the normal-tissue IsoE approximation includes the dose-independent brain weighting factors described by Pedrosa et al.~\cite{pedrosa2020isoe}.

These capabilities are not used to define the main validation endpoint of this work, which intentionally compares geometry and mesh-discretization effects using a fixed weighted-dose convention. They are nevertheless part of the TPS design because BNCT plan interpretation depends strongly on the selected biological formalism and radiobiological data. Examples of the intended applicability include beam evaluation using combined physical, radiobiological, and dosimetric figures of merit~\cite{postuma2021biology}, and comparison or combination of BNCT with carbon-ion radiotherapy through the photon-IsoE formalism~\cite{postuma2024scirep}. Future TPS-focused work will validate these model choices as separate biological-analysis layers rather than mixing them with the present geometry and dose-processing benchmark.

\subsection{Dosimetric evaluation design}
The dosimetric evaluation performed for this work was designed to test whether the open TPS processing chain preserves clinically relevant BNCT dose quantities when MCNP results are generated from a voxelized patient-style geometry rather than from an idealized analytic geometry. The comparison therefore used two representations of the same cylindrical phantom: (i) an analytic MCNP cylinder, in which the external patient boundary, shell structures, and spherical tumour were described by continuous geometric surfaces, and (ii) voxelized phantoms generated from segmented NIfTI masks using the TPS input-preparation workflow described above.

The analytic cylindrical simulation provides a controlled reference because its material boundaries are not affected by image discretization. The voxelized simulations reproduce the same nominal phantom dimensions and material classes, but represent the anatomy as a labelled lattice. This allows the study to isolate effects introduced by the TPS image-to-MCNP conversion, including voxel-size dependence, staircase approximation of curved boundaries, tally-mesh resolution, and interpolation during conversion of MCNP mesh tallies back to CT-aligned NIfTI volumes.

For each simulated case, the MCNP component tallies were converted into component-wise dose-rate maps and then combined into absorbed and weighted BNCT dose quantities using the same material labels, boron concentrations, and weighting factors. To make the comparisons independent of display resolution, dose maps and masks were evaluated on a shared reference grid. The exact analytic phantom contours were retained as the geometric reference for visual overlays, while voxelized masks were used to represent the geometry actually submitted to the TPS-style MCNP workflow.

Agreement between the analytic and voxelized simulations was assessed through three complementary endpoints. First, dose-rate quantities in the healthy brain were compared because they determine the limiting irradiation time under the selected brain constraint. Second, tissue-weighted dose distributions were compared using 3D gamma~\cite{Low1998GammaIndex,Low2003GammaEvaluation} analysis against the analytic 1 mm reference. Third, structure-wise DVH metrics were computed for GTV, brain, skull, and skin after scaling each case to the brain-limited irradiation time. This design tests the full chain from segmentation-derived material definition to MCNP input generation, dose parsing, spatial remapping, and treatment-planning metric extraction.

The tested discretization configurations are summarized in Table~\ref{tab:test-configurations}. Case names use the convention \texttt{cyl\_ms\_X\_X\_X} for analytic cylindrical simulations with an MCNP mesh size of $X$ mm, and \texttt{vs\_V\_V\_V\_ms\_M\_M\_M} for voxelized simulations with geometry voxel size $V$ mm and MCNP mesh size $M$ mm. For the voxelized sweep, only combinations with mesh size greater than or equal to the geometry voxel size were considered, so that the tally mesh was never finer than the transported lattice. All converted MCNP mesh tallies were evaluated on the common 1 mm reference support using nearest-neighbour, linear, or cubic remapping as indicated.

In this context, remapping denotes the spatial interpolation used when converting MCNP mesh-tally values to the common image grid used for NIfTI export, DVH calculation, and gamma analysis. Nearest-neighbour remapping assigns each evaluation voxel the value of the closest MCNP mesh cell and therefore preserves piecewise-constant tally values without smoothing. Linear remapping uses trilinear interpolation between neighbouring mesh-cell values, producing a continuous first-order estimate on the evaluation grid. Cubic remapping uses a higher-order cubic interpolation kernel, giving a smoother reconstructed dose field but potentially introducing larger boundary-tail differences near sharp material or dose gradients. These remapping choices affect dose reconstruction and comparison only; they do not change the underlying MCNP transport calculation.

The benchmark dataset comprised 28 independent MCNP transport simulations, each run with the equivalent of 10$^{12}$ source histories from an MCNP surface source file: six analytic cylindrical-phantom simulations with different tally-mesh sizes and 22 voxelized cylindrical-phantom simulations spanning the tested voxel-size and mesh-size combinations. The nearest-neighbour, linear, and cubic remapping comparisons were generated from these transport outputs during output processing and therefore did not require separate MCNP runs for each remapping method. The cumulative computational cost was estimated from the \texttt{computer time in mcrun} field reported in the MCNP output files. Summed over all completed transport simulations, the benchmark required 363,799 core-hours, of which 30,137 core-hours corresponded to the analytic mesh sweep and 333,662 core-hours to the voxelized-phantom sweep. Equivalently, because the simulations were run with 48 MCNP tasks, the summed task-normalized transport time was 7,579 h, or approximately 316 days, when counted sequentially over all simulations. These values represent aggregate MCNP transport time and exclude queue waiting time, failed submission attempts, and downstream Python output processing.

\begin{table}[htbp] 
\caption{Discretization configurations tested in the dosimetric evaluation. $V$ denotes the voxel size of the transported geometry and $M$ the MCNP mesh-tally size.} \label{tab:test-configurations} 
\centering 
\small 
\setlength{\tabcolsep}{5pt} \begin{tabular}{p{0.3\linewidth}p{0.26\linewidth}p{0.34\linewidth}} \toprule 
Geometry representation & Geometry voxel size $V$ [mm] & MCNP mesh-tally size $M$ [mm]\\ 
\midrule Analytic cylindrical geometry & -- & $M \in \{1,2,4,5,6,10\}\,$ \\ 
Voxelized geometry & $v \in \mathcal{V}=\{1,2,4,5,8,10\}\,$ & $M \in \{1,2,4,5,6,10\}\,$, with $M \geq v$ \\ 
Singular mesh case & 1 & 3 \\ \bottomrule \end{tabular} \end{table}

\subsection{Validation design on phantoms and patient-like geometries}
Validation is designed on two complementary scenario classes:
\begin{enumerate}
\item \textit{Synthetic phantom benchmarks}: controlled cylindrical or patient-like phantoms with known construction parameters are used to verify geometry generation, affine consistency, lattice export, and deterministic output-processing behavior.
\item \textit{Patient-like geometries}: realistic mask configurations are used to evaluate full workflow robustness (input generation, mesh parsing, CT remapping, and metric extraction) under clinically relevant spatial complexity.
\end{enumerate}

Regression tests target invariant properties (mask dimensions, affine centering, output file consistency, and reproducible component processing paths). Benchmark acceptance criteria are defined a priori through geometry consistency tests and stable dosimetric outputs across repeated runs with unchanged inputs and software versions.

For the quantitative benchmark, the analytic 1 mm cylindrical solution was used as the reference. Additional analytic cylindrical cases and voxelized phantom cases were evaluated at matched voxel and mesh resolutions, with nearest-neighbour, linear, and cubic remapping used where interpolation was required. Agreement was summarized using (i) the percent difference in brain dose-rate quantities and in the resulting brain-limited irradiation time, (ii) structure-wise gamma pass rates, and (iii) weighted DVH metric differences for tumour, brain, skull, and skin.

For each configuration, the TPS computed the biologically weighted healthy-brain dose-rate map and evaluated two dose-volume constraints: $D_2 \leq 13$ Gy-w and $D_{50} \leq 2.5$ Gy-w~\cite{pezzi2026artificial}. Here, $D_2$ is the dose received by the hottest 2\% of the healthy-brain volume, whereas $D_{50}$ is the median brain dose. The candidate irradiation times were calculated as
\begin{equation}
t_{D_2}=\frac{13\ \mathrm{Gy\mbox{-}w}}{\dot{D}_{2,\mathrm{brain}}},
\qquad
t_{D_{50}}=\frac{2.5\ \mathrm{Gy\mbox{-}w}}{\dot{D}_{50,\mathrm{brain}}},
\qquad
t_{\mathrm{brain}}=\min\left(t_{D_2},t_{D_{50}}\right).
\end{equation}
Thus, the selected treatment time was the earliest time at which either brain constraint would be reached. The $D_2$ constraint was the more restrictive one in every evaluated benchmark case; the single-voxel maximum dose, $D_{\max}$, was recorded as a diagnostic but was not used to set the irradiation time. The selected time was used to scale all structure-wise weighted DVH metrics, so that target coverage and normal-tissue endpoints were reported at the same OAR-limited treatment condition.


MCNP relative errors were first converted to absolute component-dose uncertainties, combined in quadrature after biological weighting, and propagated to DVH endpoints over the corresponding volume support: the hottest 2\% of brain voxels for $D_2$, the coldest 2\% for $D_{98}$, and a central 2\% window for $D_{50}$. When a dose mesh was remapped to a finer evaluation grid, the statistical support was expressed as the effective number of independent MCNP mesh bins in that endpoint volume, rather than the number of display voxels. The irradiation-time uncertainty was then obtained from $\sigma_t=t\,\sigma_{\dot{D}}/\dot{D}$ for the limiting brain endpoint.

\section{Results}
\subsection{Validation, regression tests, and benchmark examples}
The software-level regression tests verified the deterministic geometry and positioning operations used by the input-generation layer. The tests cover synthetic patient construction, mask dimensions, affine-centred positioning, and rotation/translation behaviour. These checks are intended to catch coordinate-system and mask-label regressions before transport calculations are launched.

The analytic cylindrical mesh sweep quantified the effect of MCNP tally-mesh size independently of voxelized-geometry effects. Table~\ref{tab:analytic-mesh-sweep} reports the percent difference in the brain $D_2$ dose rate and the resulting brain-limited irradiation time relative to the analytic 1 mm reference. Across all analytic mesh sizes and remapping methods, the absolute brain-limited time difference was at most 0.94\%. For the 2 mm analytic mesh, the time difference was 0.06\% with nearest-neighbour remapping, 0.10\% with linear remapping, and 0.07\% with cubic remapping. Coarser meshes mainly affected boundary-sensitive gamma pass rates. Patient-wide pass rates remained at least 98.10\%, and brain pass rates remained at least 99.32\%, whereas skin and tumour pass rates showed larger reductions for the coarsest meshes.

\begin{table}[htbp]
\caption{Analytic cylindrical mesh sweep against the analytic 1 mm reference. Percent differences are relative to the 1 mm analytic case processed with the corresponding evaluation workflow. Gamma criteria were 3\%/3 mm with a 10\% low-dose threshold.}
\label{tab:analytic-mesh-sweep}
\centering
\scriptsize
\setlength{\tabcolsep}{2.5pt}
\begin{tabular}{llrrrrrrr}
\toprule
Mesh & Remapping & \begin{tabular}{@{}c@{}}Time\\diff. (\%)\end{tabular} & \begin{tabular}{@{}c@{}}Brain $D_2$\\rate diff. (\%)\end{tabular} & \begin{tabular}{@{}c@{}}GTV $D_{98}$\\diff. (\%)\end{tabular} & \begin{tabular}{@{}c@{}}Skin\\$\gamma$ pass (\%)\end{tabular} & \begin{tabular}{@{}c@{}}Skull\\$\gamma$ pass (\%)\end{tabular} & \begin{tabular}{@{}c@{}}Brain\\$\gamma$ pass (\%)\end{tabular} & \begin{tabular}{@{}c@{}}GTV\\$\gamma$ pass (\%)\end{tabular} \\
\midrule
2 & nearest & 0.063 & -0.063 & 0.01 & 99.42 & 99.31 & 99.92 & 100.00 \\
4 & nearest & 0.092 & -0.092 & 3.31 & 97.54 & 98.13 & 99.77 & 99.99 \\
5 & nearest & 0.106 & -0.106 & -3.88 & 97.09 & 97.85 & 99.75 & 99.98 \\
8 & nearest & -0.005 & 0.005 & 7.92 & 96.32 & 97.51 & 99.70 & 98.55 \\
10 & nearest & -0.404 & 0.406 & -17.30 & 95.48 & 97.28 & 99.65 & 89.48 \\
2 & linear & 0.102 & -0.101 & 0.09 & 98.95 & 98.63 & 99.83 & 100.00 \\
4 & linear & 0.220 & -0.220 & 1.20 & 97.82 & 97.72 & 99.72 & 99.98 \\
5 & linear & 0.284 & -0.283 & 0.94 & 97.54 & 97.56 & 99.71 & 99.96 \\
8 & linear & 0.618 & -0.615 & 1.50 & 93.15 & 97.05 & 99.68 & 99.59 \\
10 & linear & 0.941 & -0.933 & 0.60 & 88.23 & 96.75 & 99.69 & 98.69 \\
2 & cubic & 0.070 & -0.070 & -0.03 & 99.41 & 99.15 & 99.90 & 100.00 \\
4 & cubic & 0.117 & -0.116 & 0.09 & 98.75 & 98.13 & 99.73 & 99.99 \\
5 & cubic & 0.211 & -0.211 & -0.16 & 98.37 & 97.65 & 99.69 & 99.99 \\
8 & cubic & 0.350 & -0.348 & 6.13 & 95.13 & 95.80 & 99.57 & 99.93 \\
10 & cubic & 0.176 & -0.175 & 6.24 & 90.05 & 95.90 & 99.32 & 99.82 \\
\bottomrule
\end{tabular}
\end{table}

\begin{figure}[htbp]
\centering
\includegraphics[width=\linewidth]{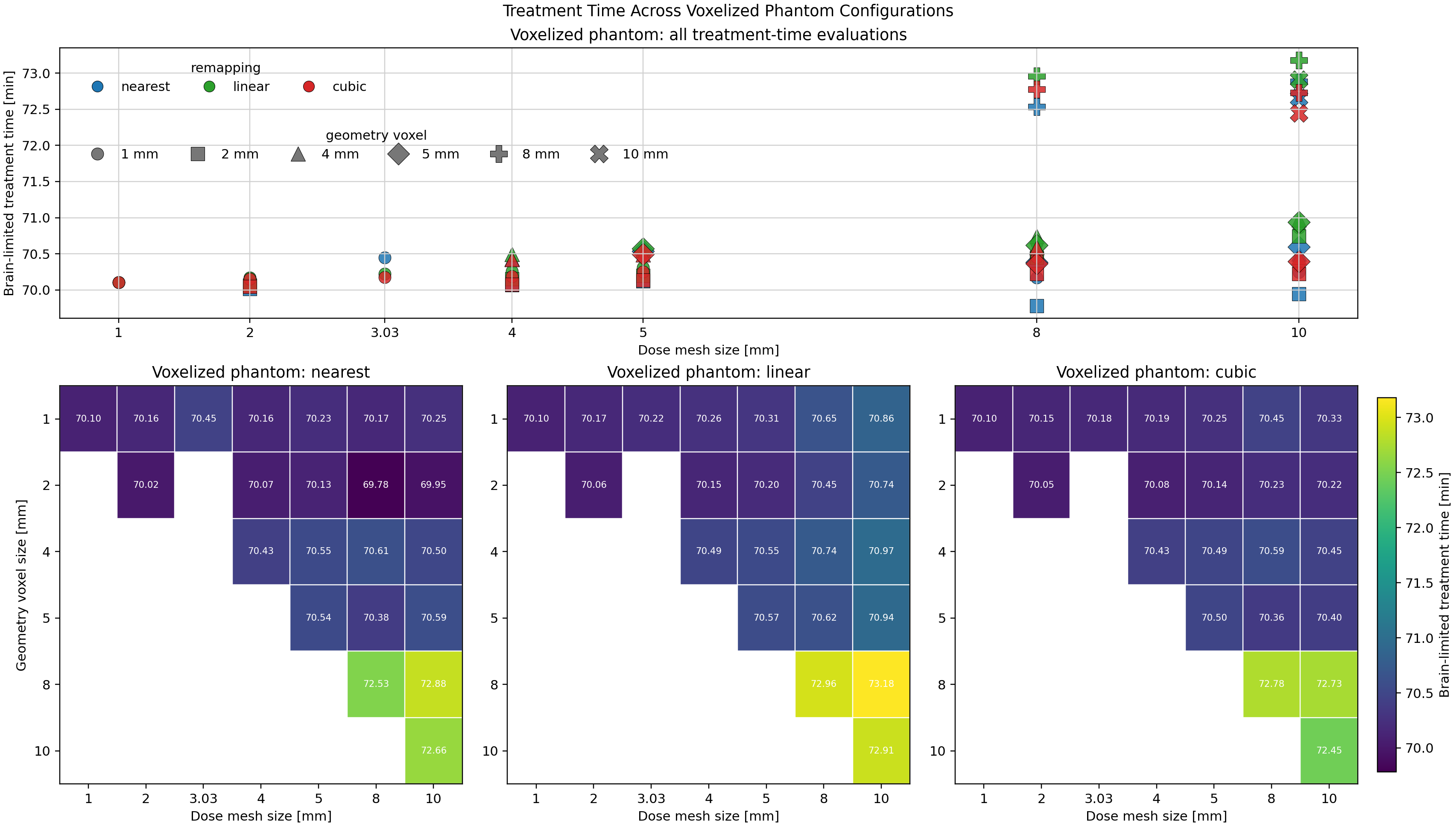}
\caption{Brain-limited irradiation time across the voxelized cylindrical-phantom sweep. The upper panel reports every voxelized configuration as a scatter plot, with point colour indicating the remapping strategy and marker shape/size indicating the geometry voxel size. The lower panels report the same results as heatmaps, with geometry voxel size on the vertical axis, MCNP tally-mesh size on the horizontal axis, and one panel for each remapping strategy. Empty heatmap cells correspond to combinations that were not simulated.}
\label{fig:treatment-time-sweep}
\end{figure}

Figure~\ref{fig:treatment-time-sweep} reports the full voxelized-phantom sweep used to assess the robustness of the voxelisation strategies. For geometry voxels and tally meshes up to \SI{5}{mm}, the brain-limited time remained tightly clustered around 70--71 min across nearest-neighbour, linear, and cubic remapping. Larger geometry voxels combined with coarse tally meshes produced longer times, reaching approximately 73 min for the 8--10 mm configurations. The largest voxelized-sweep difference relative to the analytic 1 mm nearest-neighbour reference was 4.42\%, observed for the 8 mm voxelized geometry evaluated with a 10 mm tally mesh and linear remapping. This pattern indicates that the treatment-time endpoint is stable for fine and moderate discretizations, while coarse voxelized geometries introduce a systematic change in the limiting brain $D_2$ estimate. The lower heatmaps show the same behaviour by remapping method: interpolation choice had smaller impact than the combined geometry-voxel and tally-mesh resolution, with the largest deviations occurring in the coarsest voxelized cases.

\begin{figure}[htbp]
\centering
\includegraphics[width=\linewidth]{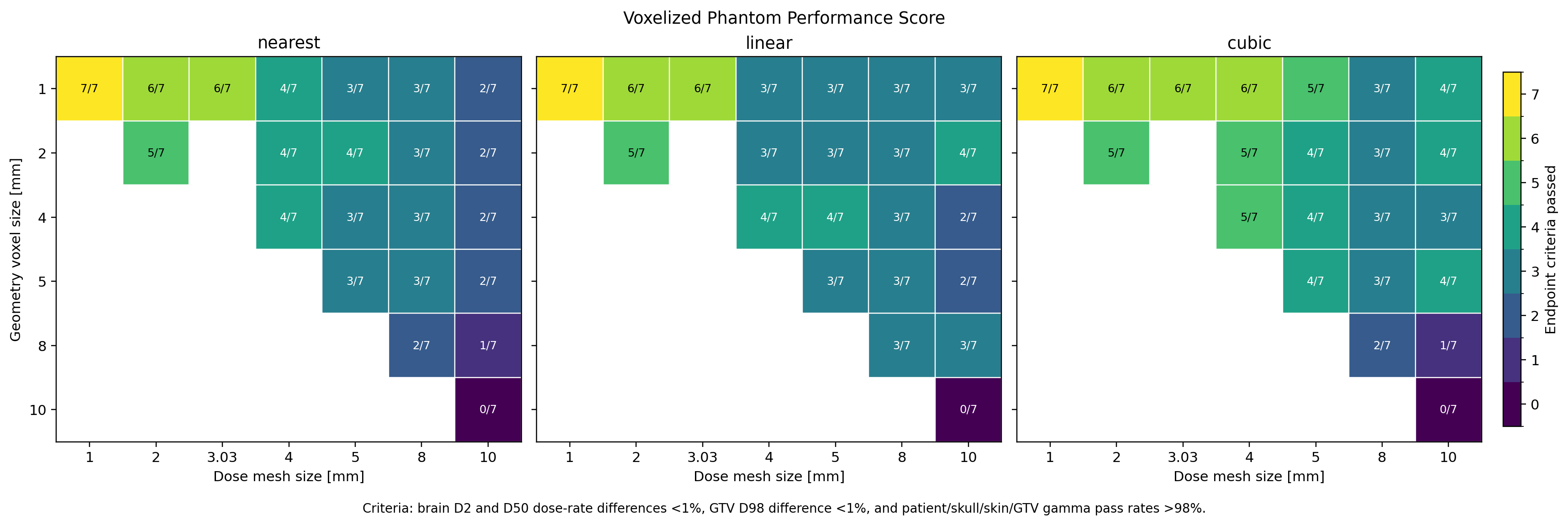}
\caption{Non-time performance score for the voxelized cylindrical-phantom sweep. Each cell reports the number of endpoint criteria satisfied out of seven. Criteria were brain $D_2$ and $D_{50}$ dose-rate differences below 1\%, GTV $D_{98}$ difference below 1\%, and patient, skull, skin, and GTV gamma pass rates above 98\%. Empty cells correspond to combinations that were not simulated.}
\label{fig:voxelized-performance-score}
\end{figure}

To summarize the non-time endpoints, each voxelized configuration was assigned a performance score from 0 to 7 (Figure~\ref{fig:voxelized-performance-score}). One point was assigned for each of the following criteria: brain $D_2$ and brain $D_{50}$ dose-rate differences below 1\%, GTV $D_{98}$ difference below 1\%, and patient, skull, skin, and GTV gamma pass rates above 98\%. The score therefore reports how many independent criteria were satisfied, rather than a mean or composite dose metric.

The highest-performing configurations were concentrated at 1 mm geometry voxel size and low tally-mesh size. The matched 1 mm geometry and 1 mm mesh satisfied all seven criteria for all three remapping methods. The 1 mm geometry with 2--3.03 mm tally meshes satisfied six of seven criteria for all remapping methods, with the remaining discrepancy driven by brain $D_{50}$ rather than by the limiting brain $D_2$ endpoint or gamma agreement. Performance decreased progressively for coarser voxelized geometries, especially for skin and skull gamma pass rates, confirming that boundary-dominated structures are the most sensitive to voxel and mesh coarsening.

Table~\ref{tab:voxelized-vs-analytic} isolates representative voxelized phantom configurations compared with the analytic 1 mm reference. The 1 mm voxelized geometry with a 1 mm tally mesh reproduced the analytic reference closely; nearest-neighbour, linear, and cubic remapping produced identical values for this case because the source mesh and evaluation grid were already matched. The 1 mm voxelized geometry with a 2 mm tally mesh and the 2 mm voxelized geometry with a 2 mm tally mesh both remained within 0.13\% for the brain-limited time endpoint. The largest DVH percent differences occurred in skull $D_{98}$, a low-dose and boundary-sensitive metric.

\begin{table}[htbp]
\caption{Voxelized phantom configurations compared with the analytic 1 mm nearest-neighbour reference. The worst structure gamma is the minimum pass rate among GTV, brain, skull, and skin. The worst DVH entry is the largest absolute weighted-DVH metric percent difference across GTV, brain, skull, and skin.}
\label{tab:voxelized-vs-analytic}
\centering
\scriptsize
\setlength{\tabcolsep}{2.2pt}
\begin{tabular}{lllrrrrrrr}
\toprule
Voxel & Mesh & Remapping & \begin{tabular}{@{}c@{}}Time\\diff. (\%)\end{tabular} & \begin{tabular}{@{}c@{}}Brain $D_2$\\rate diff. (\%)\end{tabular} & \begin{tabular}{@{}c@{}}GTV $D_{98}$\\diff. (\%)\end{tabular} & \begin{tabular}{@{}c@{}}Skin\\$\gamma$ pass (\%)\end{tabular} & \begin{tabular}{@{}c@{}}Skull\\$\gamma$ pass (\%)\end{tabular} & \begin{tabular}{@{}c@{}}Brain\\$\gamma$ pass (\%)\end{tabular} & \begin{tabular}{@{}c@{}}GTV\\$\gamma$ pass (\%)\end{tabular} \\
\midrule
1 & 1 & all & 0.029 & -0.029 & 0.20 & 100.00 & 100.00 & 100.00 & 100.00 \\
1 & 2 & nearest & 0.107 & -0.106 & 0.21 & 99.36 & 99.29 & 99.92 & 100.00 \\
1 & 2 & linear & 0.128 & -0.127 & 0.47 & 98.91 & 98.62 & 99.83 & 99.99 \\
1 & 2 & cubic & 0.097 & -0.097 & 0.25 & 99.39 & 99.11 & 99.90 & 99.99 \\
2 & 2 & nearest & -0.094 & 0.095 & 2.69 & 99.33 & 99.26 & 99.92 & 99.99 \\
2 & 2 & linear & -0.030 & 0.030 & 3.20 & 98.87 & 98.59 & 99.83 & 99.99 \\
2 & 2 & cubic & -0.051 & 0.051 & 2.90 & 99.34 & 99.10 & 99.90 & 99.99 \\
\bottomrule
\end{tabular}
\end{table}

Gamma analysis was similarly stable. Patient-wide pass rates ranged from 99.60\% to 100.00\%, with a mean of 99.85\%. The lowest structure-specific pass rates occurred in boundary-dominated normal-tissue regions, with minima of 98.59\% for skull and 98.87\% for skin. Tumour pass rates remained at or above 99.994\% in all evaluated cases. The largest observed patient mean gamma was 0.209, and the maximum gamma value was capped at 1.5 in the exported comparison summaries.

\begin{figure}[htbp]
\centering
\includegraphics[width=\linewidth]{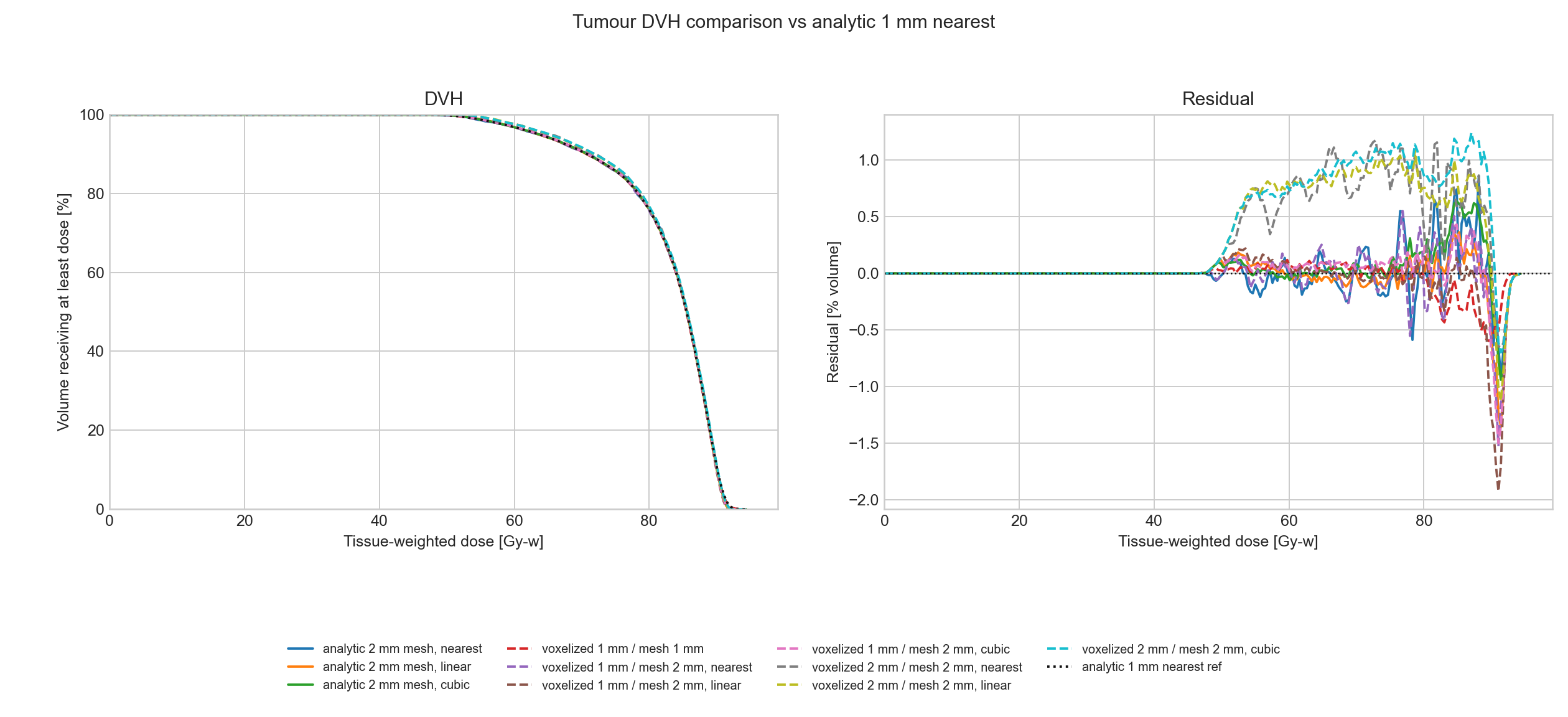}
\vspace{0.5em}
\includegraphics[width=\linewidth]{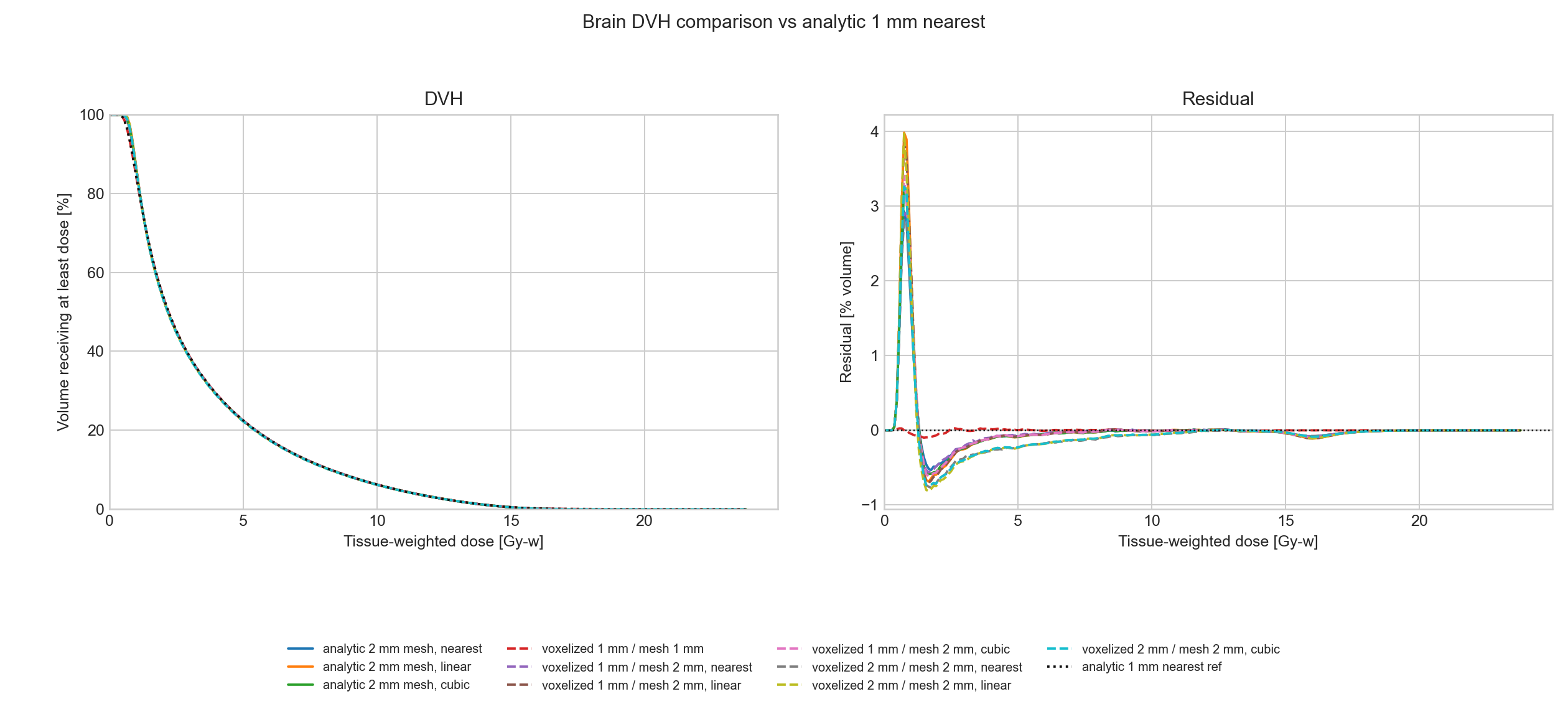}
\caption{Weighted DVH curves and DVH residuals for GTV and brain. Analytic and voxelized comparisons are shown together in each panel. The dashed black curve is the analytic 1 mm nearest-neighbour reference, and residuals are computed as candidate DVH minus reference DVH in percent volume. The analytic comparison includes the 2 mm nearest-neighbour, linear, and cubic remappings. The voxelized comparison includes the matched 1 mm voxel/1 mm mesh case once because all remappings are equivalent on the matched evaluation grid, plus the 1 mm voxel/2 mm mesh and 2 mm voxel/2 mm mesh cases with nearest-neighbour, linear, and cubic remappings.}
\label{fig:dvh-residual-target-brain}
\end{figure}

\begin{figure}[htbp]
\centering
\includegraphics[width=\linewidth]{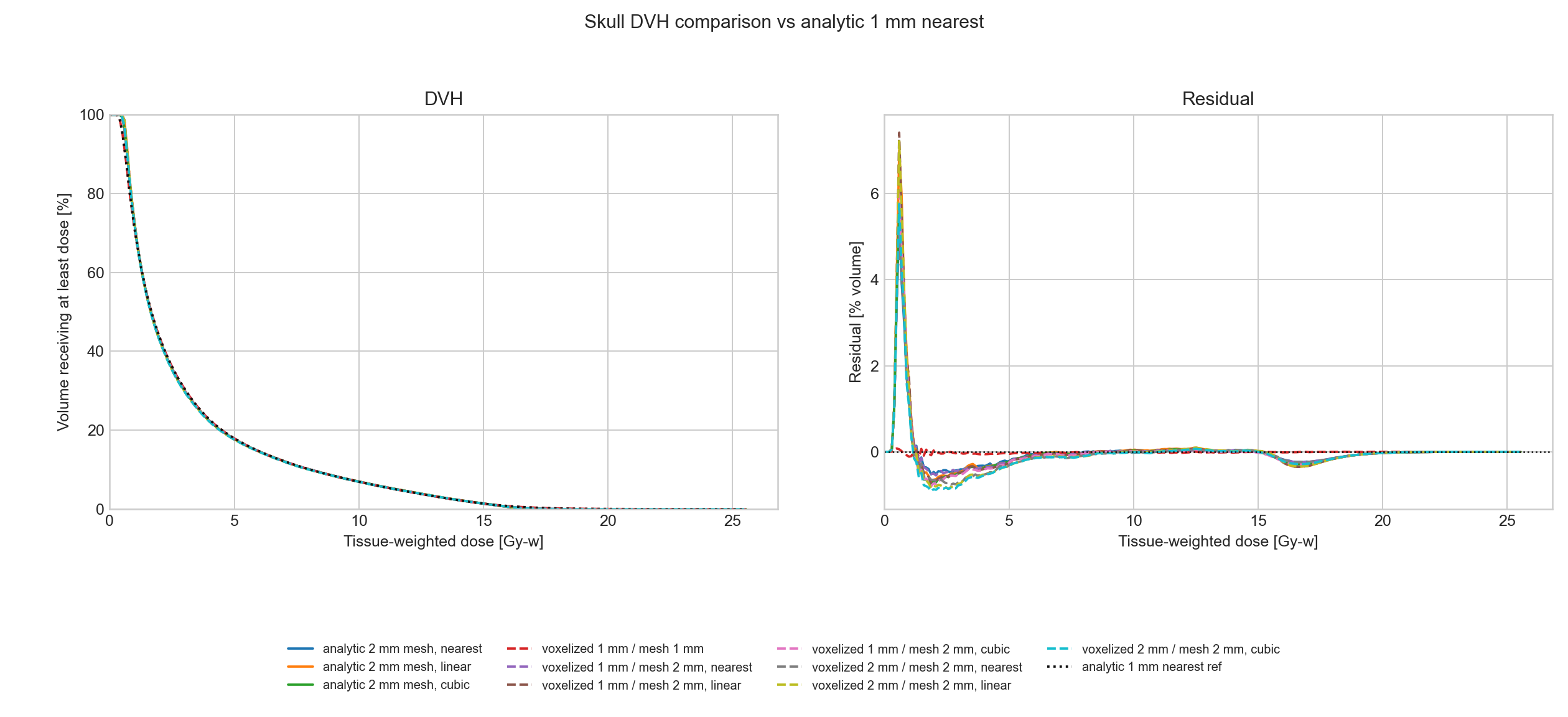}
\vspace{0.5em}
\includegraphics[width=\linewidth]{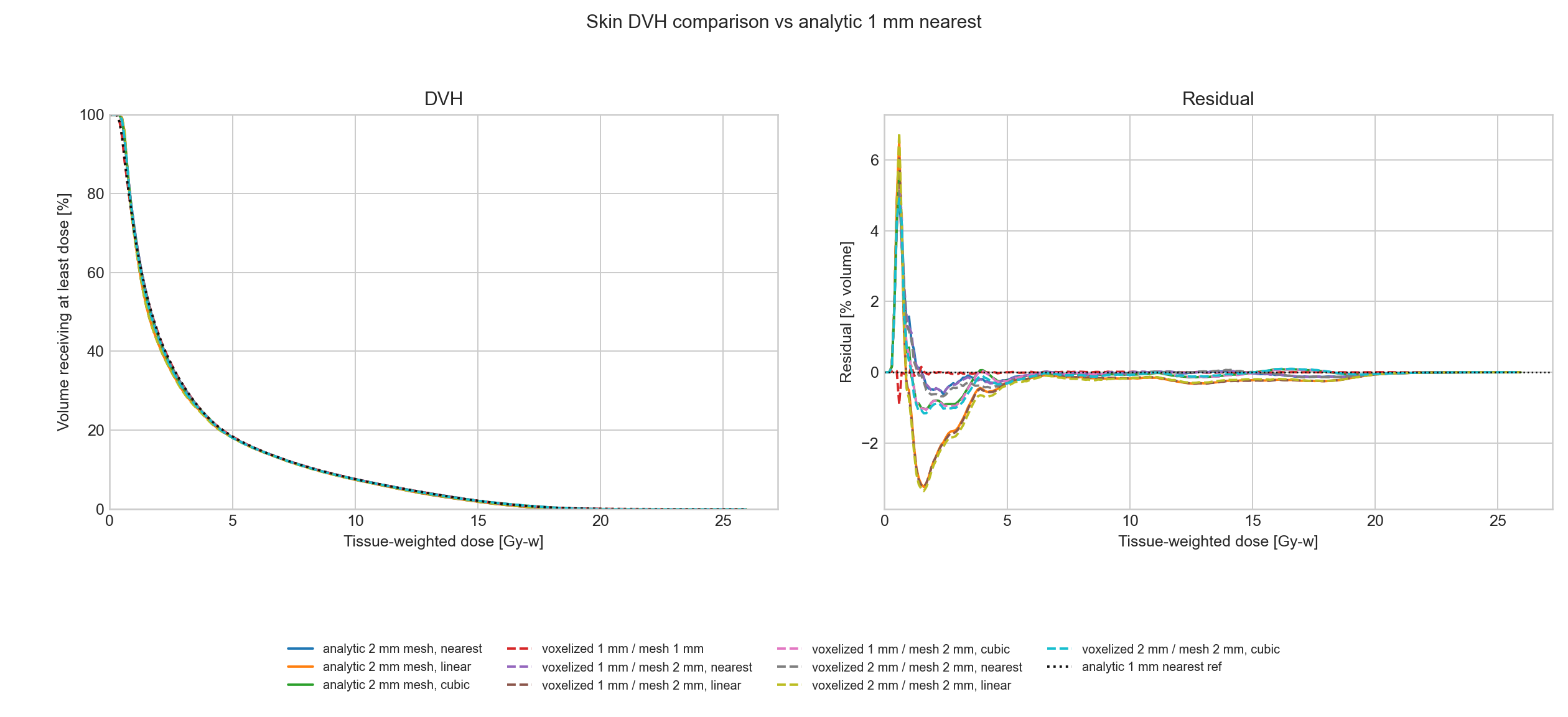}
\caption{Weighted DVH curves and DVH residuals for skull and skin. Analytic and voxelized comparisons are shown together in each panel, with residuals reported relative to the analytic 1 mm nearest-neighbour reference. The analytic comparison includes the 2 mm nearest-neighbour, linear, and cubic remappings. The voxelized comparison includes the matched 1 mm voxel/1 mm mesh case once because all remappings are equivalent on the matched evaluation grid, plus the 1 mm voxel/2 mm mesh and 2 mm voxel/2 mm mesh cases with nearest-neighbour, linear, and cubic remappings. The larger low-dose-tail residuals in these boundary-dominated normal tissues are consistent with the relative $D_{98}$ sensitivity reported in the summary metrics.}
\label{fig:dvh-residual-normal-tissues}
\end{figure}

Weighted DVH metrics were most robust for high-dose and mean-dose quantities. Across all structures and metrics, the median absolute percent difference was 0.116\%. For $D_2$ and $D_{\mathrm{mean}}$, the worst-case absolute percent differences were 1.79\% and 2.43\%, respectively. Larger relative differences were observed for $D_{98}$ in low-dose normal-tissue regions, reaching 40.85\% in skull and 34.45\% in skin. These large relative values correspond to small absolute dose levels near the lower tail of the DVH and are therefore interpreted as boundary and low-dose-volume sensitivity rather than as a failure of the high-dose planning quantities. The DVH curves and residuals are shown in Figures~\ref{fig:dvh-residual-target-brain} and~\ref{fig:dvh-residual-normal-tissues}. The residual is defined as the candidate DVH minus the analytic 1 mm nearest-neighbour reference DVH at the same tissue-weighted dose value.

\subsection{Relevance to treatment planning decisions}
The benchmark demonstrates how the module converts component-wise Monte Carlo output into quantities that can be used directly in planning review. Both the brain $D_2$ and $D_{50}$ candidate limits were evaluated for every case, and the shorter candidate time was selected. The $D_2$ constraint was reached first in all evaluated cases and produced limiting irradiation times of approximately 70--73 min, depending on mesh and interpolation choice. Because the limiting constraint was identified explicitly in the output tables, downstream planning scripts can rank candidate configurations by feasible irradiation time, target coverage at that time, and normal-tissue DVH response.


At the 1 mm reference setting, the brain-limited weighted tumour metrics were $D_{98}=57.25$ Gy-w, $D_{50}=85.55$ Gy-w, $D_2=91.46$ Gy-w, and $D_{\mathrm{mean}}=82.85$ Gy-w. In the same normalization, the brain $D_2$ was constrained to 13.0 Gy-w by construction. This illustrates the intended use of the workflow: Monte Carlo component maps are not only converted into image files, but are also reduced to traceable planning summaries that connect OAR feasibility with target-dose reporting.

\section{Discussion}
\subsection{Principal findings}
This work presents the simulation-preparation and dosimetric-analysis core of an open-source BNCT treatment-planning research platform. The workflow links MCNP input preparation, component-dose extraction, spatial remapping, and planning-oriented metric reporting. The main practical result is that the same computational path can produce both image-level outputs for inspection and table-level outputs for reproducible comparison.

The benchmark results support two conclusions. First, fine voxelized configurations reproduced the analytic reference closely for the brain-limited treatment-time endpoint, with the representative 1--2 mm configurations remaining within 0.13\%. Second, the full voxelized sweep showed that this agreement should not be generalized to arbitrarily coarse geometries: 8--10 mm voxel/mesh combinations shifted the limiting time by up to 4.42\%, and low-dose DVH-tail metrics in thin normal tissues were especially sensitive to boundary discretization. Gamma analysis provided a compact 3D agreement summary, but it was interpreted together with DVH endpoints rather than as a substitute for structure-specific dose review.

The separation between physical component maps and weighted dose summaries is important for auditability. Component maps retain the transport-derived information needed for independent review, whereas weighted maps and DVH metrics encode the assumptions used for planning interpretation. This design makes it possible to repeat the same transport-output processing with alternative weighting factors or tissue boron concentrations without regenerating the MC input or changing the geometry pipeline.

\subsection{Limitations}
The present manuscript validates the computational core of an open-source treatment-planning platform rather than a complete clinical treatment planning system. The quantitative dose-agreement benchmark is based on controlled cylindrical phantom geometries; the patient-positioning example demonstrates geometric workflow integration but does not constitute patient-cohort dosimetric validation. Consequently, the reported agreement should be interpreted as evidence of implementation consistency and numerical stability, not as evidence of clinical outcome prediction.

The current workflow also assumes that the input masks, material assignments, source description, and biological weighting parameters supplied by the user are appropriate for the intended study. It does not calibrate boron pharmacokinetics, optimize beam arrangements, or estimate patient-specific biological response. These tasks require additional modelling layers and clinical validation beyond the scope of this standalone module.

The current kerma-factor handling is also intentionally conservative. The benchmark uses reference BNCT kerma tables rather than generating composition-specific kerma factors for each material definition. This is appropriate for validating the present processing chain against a fixed reference convention, but it limits material specificity in heterogeneous patient models. A more complete research TPS should support material contouring, material-composition management, and kerma-factor generation from elemental or isotopic composition.

Finally, interpolation and discretization effects remain visible in low-dose DVH-tail quantities, especially $D_{98}$ in thin or boundary-dominated normal tissues. For treatment-decision endpoints that depend on low-dose tails, users should inspect the structure geometry, mesh resolution, interpolation method, and propagated MCNP endpoint uncertainty rather than relying only on percent differences.

\subsection{Roadmap toward the full BNCT TPS}
The next development stage is expansion of this validated computational core into a broader BNCT TPS framework. That framework will add optimization strategies, expanded biological weighting models, plan-comparison interfaces, and a larger validation set including patient-specific case studies. In parallel, coupling \texttt{OpenPINT.mcgenerator} with the \texttt{mcgeometry} framework will improve the consistency of geometric modelling, source placement, and simulation-domain construction~\cite{mcgeometry_zenodo}.

A specific missing capability is automated, or AI-assisted, material contouring at the beginning of the planning workflow. Such a system could help assign material classes more consistently and would support the use of kerma factors specific to the different material compositions present in the patient model. In future releases, a complementary kerma-generation feature should derive material-specific kerma factors directly from the defined elemental or isotopic composition, reducing reliance on generic material assumptions and improving the physical specificity of the dose calculation.

Subsequent papers will therefore focus on treatment-planning behaviour rather than on the lower-level simulation-preparation and dose-analysis mechanics isolated here. Keeping these contributions separate allows the present release to be cited as the reproducible MCNP-input and dosimetric-output layer, while later TPS manuscripts can address optimization, user workflow, and clinical interpretation.

\section{Code availability, version DOI, license, and reproducibility statement}
The source code is distributed as the open-source \texttt{OpenPINT} repository~\cite{bnctdosimetry} under the European Union Public Licence v1.2 (EUPL-1.2). The public repository is available at \url{https://github.com/ipostuma/OpenPINT}. For submission, the manuscript results should be tied to a fixed release tag or commit and to an archived software DOI. Dependencies are listed in \texttt{requirements.txt}, and executable workflow entry points are provided in \texttt{bin/}.

The benchmark scripts used for this manuscript are stored under  \texttt{examples/ mc\_input\_generator/ pub\_sim/}. All supplementary examples, benchmark inputs, generated masks, patient-positioning outputs, figures, validation tables, and simulated comparison data for the voxel-size and mesh-size study are archived in a \href{\restrictedzenodolink}{restricted Zenodo data record}~\cite{openpint_supplementary_zenodo}. The final reference-comparison summaries are exported as CSV files containing brain-constraint scaling, gamma pass rates, and weighted DVH metric differences. These files provide the numerical source for Figures~\ref{fig:treatment-time-sweep} and~\ref{fig:voxelized-performance-score}, and allow the reported values to be regenerated from the repository scripts and archived data.

\section{Conclusion}
This work presents the reproducible MCNP simulation-preparation and dosimetric-analysis core of an open-source BNCT treatment-planning research platform, with validation-oriented design choices that support reliable planning studies. By exposing the full path from segmented images and simulation inputs to component dose maps, biological weighting, DVH metrics, gamma analysis, and archived benchmark tables, the repository provides more than a collection of scripts: it provides a common technical basis on which BNCT methods can be tested, compared, and extended.

This openness can accelerate BNCT research by reducing duplicated implementation work and by making numerical assumptions visible. Researchers developing beam-shaping assemblies, source models, geometry tools, biological weighting models, boron-distribution methods, imaging workflows, or plan-evaluation metrics can use the same repository as a shared reference layer. This makes it easier to isolate whether differences between studies arise from transport physics, patient modelling, biological assumptions, interpolation, mesh resolution, or treatment-planning criteria. In this way, the software can help connect groups working on different BNCT aspects and turn independent developments into comparable contributions. A clear example is the evaluation of different beams performances, where spectral characteristics, gamma contamination, geometrical aspects can play a role in the effectiveness of the treatment. The use of a shared computational tool for this kind of comparisons adds to the robustness of evaluations across the facilities. 

The present release should therefore be viewed as a reproducible foundation for a broader community-facing BNCT TPS framework. Future developments will expand optimization, biological-analysis, patient-specific validation, material-contouring, and composition-specific kerma-generation capabilities, while coupling \texttt{OpenPINT.mcgenerator} to the \texttt{mcgeometry} framework will strengthen geometric modelling integration and end-to-end workflow consistency~\cite{mcgeometry_zenodo}. The long-term aim is for the repository to support not only local planning studies, but also multi-institutional benchmarking, method validation, and collaborative development of transparent BNCT treatment-planning tools.


\funding{This work was funded by INFN National Scientific Committee 5 Young Grant IT\_STARTS. It was also supported by the National Plan for NRRP Complementary Investments (PNC, established with decree-law 6 May 2021, n. 59, converted by law n. 101 of 2021) in the call for the funding of research initiatives for technologies and innovative trajectories in the health and care sectors (Directorial Decree n. 931 of 06-06-2022), project n. PNC0000003, AdvaNced Technologies for Human-centrEd Medicine (ANTHEM). This work reflects only the authors' views and opinions; neither the Ministry for University and Research nor the European Commission can be considered responsible for them.}

\roles{I.P.: Conceptualization, software, methodology, validation, writing---original draft, writing---review and editing.
       S.J.G.: Conceptualization, methodology, writing---original draft, writing---review and editing.  
       S.F.: Validation, writing---review and editing.
       C.P.: Validation.
       C.R.: Methodology, validation.
       O.N.: Conceptualization, writing---original draft, writing---review and editing.
       V.V.: Conceptualization, writing---original draft, writing---review and editing.
       S.B.: Conceptualization, methodology, validation, writing---original draft, writing---review and editing.}

\data{The \texttt{OpenPINT} source code is available at \url{https://github.com/ipostuma/OpenPINT}. All supplementary examples, benchmark data, generated masks, patient-positioning outputs, figures, and validation tables are archived in a \href{\restrictedzenodolink}{restricted Zenodo data record}~\cite{openpint_supplementary_zenodo}.}

\suppdata{Supplementary material consists of the full OpenPINT \texttt{examples/} directory, a Zenodo README, benchmark configuration files, generated masks, patient-positioning example outputs, supplementary figures, and extended validation tables associated with the archived software and simulation-data records in the \href{\restrictedzenodolink}{restricted Zenodo data record}~\cite{openpint_supplementary_zenodo}.}

\section*{References}
\bibliographystyle{unsrt}
\bibliography{references}

\end{document}